\documentclass[journal]{IEEEtran}
\usepackage{amsmath,amssymb,verbatim,cite,citesort,amsopn,graphicx,url,color}
\usepackage{algorithm}
\usepackage{multicol}
\usepackage{epsfig}
\usepackage{epstopdf}
\usepackage{subcaption}
\usepackage{balance}
\usepackage[font={small}]{caption}
\usepackage{enumerate}
\usepackage{algpseudocode}
\usepackage{hhline}%

\newcommand{\ry}{\mathrm{ry}}
\newcommand{\cl}{\mathrm{cl}}

\newcommand{\out}{\mathrm{out}}
\newcommand{\PP}{\mathbb{P}}
\newcommand{\EE}{\mathbb{E}}

\newcommand{\sgn}{\mathrm{sgn}}
\newcommand{\erf}{\mathrm{erf}}

\newcommand{\DC}{\mathbb{C}}

\newcommand{\HV}{\mathbf{H}_{\psi,V}}
\newcommand{\sHV}{\widetilde{\mathbf{H}}_{\psi,V}}
\newcommand{\sxV}{\widetilde{\mathbf{x}}_{V}}
\newcommand{\syV}{\widetilde{\mathbf{y}}_{V}}
\newcommand{\snV}{\widetilde{\mathbf{n}}_{V}}

\newcommand{\cssHV}{\widehat{\widetilde{\mathbf{H}}}_{\psi,V}}
\newcommand{\cpsi}{\widehat{\psi}}
\newcommand{\cHV}{\mathbf{H}_{\widehat{\psi},V}}
\newcommand{\csHV}{\widehat{\widetilde{\mathbf{H}}}_{\widehat{\psi},V}}
\newcommand{\ssHV}{\widetilde{\mathbf{H}}_{\widehat{\psi},V}}
\newcommand{\cbHV}{\widehat{\widetilde{\mathbf{H}}}_{\psi,V}}

\newcommand{\Mset}{\mathcal{M}}
\newcommand{\MsetR}{\mathcal{M}_{r}}
\newcommand{\MsetT}{\mathcal{M}_{t}}

\newcommand{\ba}{\mathbf{a}}
\newcommand{\bA}{\mathbf{A}}
\newcommand{\bF}{\mathbf{F}}

\newcommand{\bH}{\mathbf{H}}
\newcommand{\bh}{\mathbf{h}}
\newcommand{\bJ}{\mathbf{J}}
\newcommand{\bM}{\mathbf{M}}
\newcommand{\bn}{\mathbf{n}}

\newcommand{\br}{\mathbf{r}}
\newcommand{\bs}{\mathbf{s}}

\newcommand{\bX}{\mathbf{X}}
\newcommand{\bx}{\mathbf{x}}
\newcommand{\by}{\mathbf{y}}

\newcommand{\bI}{\mathbf{I}}
\newcommand{\eps}{\epsilon}

\newcommand{\Tr}{\mathrm{Tr}}

\newcommand{\varR}{\sigma^2_{R_\psi}}
\newcommand{\muR}{\bar{R}_{\psi}}

\newtheorem{Proposition}{Proposition}

\newtheorem{Corollary}{Corollary}

\begin{document}

\title{Low-Complexity Reconfigurable MIMO for Millimeter Wave Communications}

\author{
Biao~He,~\IEEEmembership{Member,~IEEE,}
and Hamid Jafarkhani,~\IEEEmembership{Fellow,~IEEE}
\thanks{This work was presented in part at the 2018 IEEE International Conference on Communications (ICC)~\cite{He_18_mwcwras}.}
\thanks{This work was supported in part by the NSF Award ECCS-1642536.}
\thanks{The authors are with the Center for Pervasive Communications and Computing,
University of California at Irvine, Irvine, CA 92697, USA (email:
\{biao.he, hamidj\}@uci.edu).}
}

\maketitle
\begin{abstract}
The performance of millimeter wave (mmWave) multiple-input multiple-output (MIMO) systems is limited by the sparse nature of propagation channels and the restricted number of radio frequency chains at transceivers.
The introduction of reconfigurable antennas offers an additional degree of freedom on designing mmWave MIMO systems. This paper provides a theoretical framework for studying the mmWave MIMO with reconfigurable antennas. Based on the virtual channel model, we present an architecture of reconfigurable mmWave MIMO with beamspace hybrid analog-digital beamformers and reconfigurable antennas at both the transmitter and the receiver. We show that employing reconfigurable antennas can provide  throughput gain for the mmWave MIMO. We derive the expression for the average throughput gain of using reconfigurable antennas in the system, and further derive the expression for the outage throughput gain  for the scenarios where the channels are (quasi) static.
Moreover, we propose a low-complexity algorithm for reconfiguration state selection and  beam selection.
Our numerical results verify the derived expressions for the throughput gains  and demonstrate the near-optimal throughput performance of the proposed low-complexity algorithm.
\end{abstract}

\begin{IEEEkeywords}
Reconfigurable antennas, millimeter wave, sparse channels, virtual channel model, throughput gain.\vspace{-5mm}
\end{IEEEkeywords}

\section{Introduction}
The fast development of wireless technology has enabled the  ubiquitous  use of wireless devices in modern life.
Consequently, a capacity crisis in wireless communications is being created.
A promising solution to the capacity crisis is to increase the available spectrum for commercial wireless networks by exploring the millimeter wave (mmWave) band from 30 GHz to 300 GHz.

MmWave communications has drawn significant attention in recent years~\cite{rappaport2014millimeter}, as the large available bandwidth may offer multiple-Gbps data rates~\cite{Pi11Aninmmvmbs}.
In particular, the 3rd Generation Partnership Project (3GPP), which is a work in progress to serve as the international industry standard for 5G, has continuously published the technical report (TR) documents on the  mmWave  channels, e.g.,~\cite{3GPPTR389011411,3GPPTR389001420,3GPPR1-164975}. 
Different from low-frequency communications, mmWave carrier frequencies are relatively very high. The high frequency results in some propagation challenges, such as large pathloss and severe shadowing~\cite{rappaport2014millimeter}.
Meanwhile, the small wavelength enables a large number of antennas to be closely packed to form mmWave large multi-antenna systems, which can be utilized to overcome the propagation challenges and provide reasonable signal to noise ratios (SNRs)~\cite{rappaport2013millimeter}.
The multi-input multi-output (MIMO) technology has already been standardized and widely adopted in commercial WLAN and cellular systems at sub-6 GHz frequencies (IEEE 802.11n/ac, IEEE 802.16e/m, 3GPP cellular LTE, and LTE Advanced)~\cite{Kim7060495,Li5458368}.
However, the performance of mmWave MIMO systems is still considerably limited due to the sparsity of the channels and the stringent constraint of using radio frequency (RF) chains in mmWave transceivers.
The directional propagations and clustered scattering make the mmWave paths to be highly sparse~\cite{Pi11Aninmmvmbs}. More importantly, the high cost and power consumption of RF components and data converters preclude the adoption of fully digital processing for mmWave MIMO to achieve large beamforming gains~\cite{Pi11Aninmmvmbs,Doan04dcf60gcmosra}, and low-complexity transceivers relying heavily on analog or hybrid (analog-digital) processing are often adopted~\cite{Venkateswaran10ancpsnoce,Ayach_14_SpatiallySparsePrecodingmmMIMO,Liu06STTrecbocpf}.

The limited  beamforming capability and performance of mmWave MIMO motivate us to investigate the  potential benefits of employing reconfigurable antennas for mmWave MIMO in this work.
Different from conventional antennas with fixed radiation characteristics,
reconfigurable antennas can dynamically change their radiation patterns, polarizations, and/or frequencies~\cite{Cetiner_04_MEMS_Magazine,Grau_08_AreMIMOcym}, and offer an additional degree of freedom for designing mmWave MIMO systems.
The radiation characteristics of an antenna is directly determined by the distribution of its current~\cite{balanis2005antenna}, and the mechanism of reconfigurable antennas is to change the current flow in the antenna,
so that the radiation pattern, polarization, and/or frequency can be modified. 
The detailed
relationship between  the geometry of an antenna's current and
how it radiates or collects the energy can be found in Equation (1) in \cite{Grau_08_AreMIMOcym}.
The study of reconfigurable antennas for traditional low-frequency MIMO has received considerable attention, e.g., presented in~\cite{Cetiner_04_MEMS_Magazine,Grau_08_AreMIMOcym,Fazel09stsbcmmioyra,Christodoulou12Rafwsa,Haupt13ra,Pendharker14ocfrmalp}. From the practical antenna design perspective, different approaches to make antennas reconfigurable have been proposed and realized, such as Microelectrophoretical Systems (MEMS) switches,  diodes, field-effect transistors (FETs), varactors, and optical switches~\cite{Christodoulou12Rafwsa,Haupt13ra,Pendharker14ocfrmalp}.
From the perspective of theoretical performance analysis, the array gain, diversity gain, and coding gain~\cite{Jafarkhani_05_stctpbook} of employing reconfigurable antennas have been derived with space-time code designs~\cite{Grau_08_AreMIMOcym,Fazel09stsbcmmioyra}.
More recently, reconfigurable antennas for communications at mmWave frequencies have been designed and realized, e.g., \cite{Jilani_16_FMMFRdsds} at 20.7--36 GHz,  \cite{Ghassemiparvin_16A_Rmmsdedfs} at 92.6–-99.3 GHz, and~\cite{Costa_17_OpticallCRmmAas} at 28--38 GHz.
The design of space-time codes  for a $2\times2$ mmWave MIMO with reconfigurable transmit antennas was investigated in~\cite{Vakilian_15_SThmmra} and~\cite{Vakilian_15_ThmmraAr}, and the achieved diversity gain and coding gain were demonstrated.
Due to the simple structure of a $2\times2$ MIMO, neither the important sparse nature of mmWave channels nor the transceivers with low-complexity beamforming were considered in~\cite{Vakilian_15_SThmmra} and~\cite{Vakilian_15_ThmmraAr}.


In this work, we comprehensively study the mmWave MIMO systems with reconfigurable antennas.
We consider that the radiation patterns and/or the polarizations of antennas are reconfigurable, while do not consider the frequency reconfigurations.
We consider general mmWave systems over sparse channels and the transceivers with low-complexity beamforming are taken into account.
Our analytical results are generally applicable for the whole mmWave frequency range where the channel matrices are sparse provided the models are reflective of real systems.
We present a practical architecture of mmWave MIMO with beamspace hybrid beamformers and reconfigurable antennas.
The presented architecture has a low-complexity structure for practical implementation, since it only requires a few RF chains. More importantly, as will be shown later in the paper, the presented architecture offers tractable analytical results on the throughput gains of employing reconfigurable antennas.
The throughput gains  of employing reconfigurable antennas for the mmWave system are investigated, and a fast selection algorithm for the antennas' reconfiguration state and the hybrid beamformers' beams  is further proposed.

The primary contributions of the paper are summarized as follows:
\begin{enumerate}
  \item We are the first to provide a theoretical framework for studying the reconfigurable antennas in mmWave MIMO systems. We take the sparse nature of mmWave channels into account, and present a practical architecture of the mmWave MIMO with low-complexity beamformers and reconfigurable antennas, in which beamspace hybrid beamformers and  reconfigurable antennas are employed at both the transmitter and the receiver. 
  \item We investigate the throughput gains of employing reconfigurable antennas in mmWave MIMO systems. We derive the expression for the average throughput gain, which involves an infinite integral of the error function. We further consider the cases of small and large  numbers of reconfiguration states, and derive the corresponding simplified expressions for the average throughput gains. We also derive the expression for the outage throughput gain of employing reconfigurable antennas for the (quasi) static channels. Moreover, we analyze the limiting growth rates of the average throughput gain and the outage throughput gain as the number of reconfiguration states becomes large.
      To the best of our knowledge, the throughput gains of employing reconfigurable antennas have never been derived in the literature, even in the case of  low-frequency systems.
  \item
  With the highly sparse nature of the channel, the number of non-vanishing rows and columns of the channel matrix in beamspace domain is relatively small,  and the dominant beams usually significantly outperform the others.
  Taking those advantages of the sparse nature of mmWave channels, we propose a fast algorithm for selecting the reconfiguration state of the antennas and the beams for the beamspace hybrid beamformers.
  The proposed algorithm significantly reduces the complexity of the reconfiguration
  state selection and beam selection without a large throughput loss compared with the optimal selection of reconfiguration state and beams by exhaustive search.
  \item We demonstrate the throughput gains of employing reconfigurable antennas in mmWave MIMO systems by numerical evaluations. A practical clustered multipath model is adopted for generating MIMO channels~\cite{akdeniz2014millimeter}. Our results show that the employment of reconfigurable antennas provides both the average throughput gain and the outage throughput gain for mmWave MIMO systems, and confirm the accuracy of our derived expressions.
      For the outage throughput, an interesting finding is that the performance gain increases as the required outage level becomes more stringent.
\end{enumerate}

The notations are summarized in Table~\ref{table:Notations}.
\begin{table}[t]
\centering
\caption{Summary of notations.}  \label{table:Notations}
\begin{tabular}{|c|p{6cm}|}
\hline
symbol & meaning \\
 \hhline{|=|=|}
 $\bX^T$ & transpose of $\bX$ \\
\hline
 $\bX^H$ & conjugate transpose of $\bX$ \\
 \hline
 $\bX\left(m,n\right)$  &  entry of $\bX$ in $m$-th row and $n$-th column \\
 \hline
 $\Tr(\bX)$  & trace of $\bX$ \\
 \hline
 $\left|\bX\right|$  &  determinant of $\bX$ \\
 \hline
 $\left\|\bX\right\|_F$ & Frobenius norm of $\bX$ \\
 \hline
 $\mathrm{Re}[x]$ & real part of $x$ \\
 \hline
 $\mathrm{Im}[x]$    &  imaginary part of $x$ \\
 \hline
  $\odot$  & Hadamard (element-wise) product \\
  \hline
 $\left|\mathcal{X}\right|$  & cardinality of set $\mathcal{X}$ \\
 \hline
 $\sgn(\cdot)$   & sign function \\
   \hline
 $\mathrm{erf}(\cdot)$  &   error function \\
   \hline
$\mathrm{erf}^{-1}(\cdot)$   &  inverse error function \\
   \hline
$\EE\{\cdot\}$    &  expectation operation \\
   \hline
$\PP(\cdot)$    &  probability measure \\
   \hline
$\mathrm{corr}(\cdot,\cdot) $ & correlation coefficient \\
   \hline
 $\bI_n$  &  identity matrix of size~$n$ \\
   \hline
 $\mathcal{N}(\mu,\sigma^2)$   &  Gaussian distribution with mean $\mu$ and variance $\sigma^2$ \\
   \hline
$\mathcal{CN}(\mu,\sigma^2)$   & complex Gaussian distribution with mean $\mu$ and variance $\sigma^2$ \\
   \hline
$\mathcal{CN}(\ba,\bA)$   &  distribution of a circularly symmetric complex Gaussian random vector with mean $\ba$ and covariance matrix $\bA$\\
   \hline
\end{tabular}
\end{table}

\section{Preliminaries}

Different from conventional antennas with fixed radiation characteristics, reconfigurable antennas can dynamically change their radiation characteristics.
The mechanism of reconfigurable antennas is to change the current flow in the antenna, so that the radiation pattern, polarization, and/or frequency can be modified.
Specifically, the manner how an antenna radiates or captures the energy can be mathematically expressed as~\cite{balanis2005antenna,Grau_08_AreMIMOcym}
\begin{equation}\label{eq:radiantennafo}
  \bF(\theta,\phi)=\hat{\br}\times\left[\hat{\br}\times\left[\int_{V'}\bJ_{V'(\br')}e^{-j\beta\hat{\br}\br}\mathrm{d}v'\right]\right],
\end{equation}
where $\bF(\theta,\phi) = \left(F_{\theta}(\theta,\phi), F_{\phi}(\theta,\phi)\right)^\dagger$  denotes the normalized complex amplitude radiation pattern of the antenna,
$\bJ_{V'(\br')}$ denotes the current distribution in the antenna,
$\br'$ denotes the vector from the coordinate system's origin to any point on the antenna,
$\hat{\br}$ denotes the unit vector in the direction of propagation (from the origin to the observation point),
$\beta$ denotes the propagation constant,
$V'$ denotes the volume of the antenna
containing the volumetric current densities, and $\theta$ and $\phi$ denote
the elevation and azimuth angles in the spherical coordinate
system, respectively.
It is evident from \eqref{eq:radiantennafo} that one can
change the current distribution $\bJ_{V'(\br')}$ by
altering the antenna's physical configuration $\br'$, and finally modify the antenna's radiation
characteristics $\bF(\theta,\phi)$. There are different approaches to change the antenna's  physical configuration in practice, e.g., microelectrophoretical systems (MEMS),  diodes, field-effect transistors (FETs), varactors, and optical switches~\cite{Christodoulou12Rafwsa,Haupt13ra,Pendharker14ocfrmalp}.
One  can  find  examples  of  how  the  change  in  current  distribution  affects  the  antenna's  radiation  pattern,  polarization,  and  frequency  in~\cite{Cetiner_04_MEMS_Magazine}.
 The design of reconfigurable antennas for mmWave frequencies has been specifically investigated in, e.g.,~\cite{Jilani_16_FMMFRdsds,Ghassemiparvin_16A_Rmmsdedfs,Costa_17_OpticallCRmmAas}.
It is worth mentioning that the concept of reconfigurable antennas in this paper is different from the concept of reconfigurable antenna arrays in, e.g.,~\cite{Sayeed_07_maxMcsparseRAA}.
As previously explained, the reconfigurable antennas are realized by changing the current flow  in the antenna, while the reconfigurable antenna arrays in, e.g.,~\cite{Sayeed_07_maxMcsparseRAA}, are achieved by changing array configurations (antenna spacings).

Although existing studies have shown that mmWave reconfigurable antennas can be realized~\cite{Jilani_16_FMMFRdsds,Ghassemiparvin_16A_Rmmsdedfs,Costa_17_OpticallCRmmAas}, there still exist a number of challenges in practical implementation. In the following, we specifically discuss four issues of reconfigurable antennas that one may encounter in practical implementation, which are (1) switching delay, (2) power consumption, (3) form factor, and (4) cost.
A major issue of implementing reconfigurable antennas in practice for not only sub-6 GHz systems but also mmWave communications is the switching delay incurred by changing the reconfiguration states. In practical implementation, switching the reconfigurable antennas from one state to another takes non-negligible switching time~\cite{Fazel09stsbcmmioyra}. For example, a switch device with a switching time of 100 microsecond can cause considerable delays for mmWave systems.
Another issue for the implementation of reconfigurable antennas is the power consumption. Compared with the traditional antennas, the reconfigurable antennas require an extra power consumption to enable the switches for reconfiguration. This issue becomes more critical considering the high power consumption of mmWave circuits and systems. 
The third issue for the implementation of reconfigurable antennas is their relatively large size. As the wavelength becomes small, mmWave transceivers can pack a large number of antennas in a compact size to overcome the propagation challenges. However, the employment of switches for reconfiguration  makes the size of reconfigurable antennas relatively larger than traditional antennas. Thus, packing a large number of reconfigurable antennas together may result in a large form factor.
Last but not least, the additional cost of circuits and switches to operate and control the antenna reconfiguration is also an issue for the implementation. In particular, the requirement of low cost would make it more challenging to address the aforementioned three issues in practical implementation.

To address the implementation issues of reconfigurable antennas, considerable research efforts on the circuit and switch designs are still needed, since the switching time, power consumption, form factor, and cost all highly depend on specific circuit topologies and switch technologies.
For example, the pin diodes and MEMS switches have a relatively low power consumption and fast switching speed. However, these devices experience larger losses at high frequencies.
A possible solution is to explore smart material based switches,  via metal-insulator transition compounds such as vanadium oxide ($VO_2$), which are designed to operate at mmWave frequencies~\cite{Khalat10259653}.
One advantage is that smart materials do not require energy to maintain either the ON (crystalline state) or OFF (amorphous state) state, which thus reduces power consumption in switching the reconfiguration states. Also, relatively fast switching speed may be achievable by $VO_2$ based switches, where the demonstrated fastest switching speed is about 5 nanosecond~\cite{Zhou6403505}.

\section{System Model}\label{sec:sysmod}
We consider a mmWave system where a transmitter with $N_t$ antennas sends messages to a receiver with $N_r$ antennas.
The antennas at both the transmitter and the receiver are reconfigurable.
We assume that the transmit antennas can be reconfigured into $Q$ distinct radiation states and the receive antennas can be reconfigured into $W$ distinct radiation states.
Here two radiation states are distinct when the antenna's radiation characteristics associated with the two states are orthogonal to each other. From an electromagnetic point of view, orthogonal radiation characteristics can be generated by using polarization, pattern, space, or frequency diversity techniques, or any combination of  them~\cite{Waldschmidt1321326,Grau_08_AreMIMOcym,Huff1589416,Aissat1643626}.
For example, polarization diversity imposes orthogonality by creating  orthogonal polarizations and pattern diversity imposes orthogonality by producing spatially disjoint radiation patterns.
From \eqref{eq:radiantennafo},  we note that one can modify the antenna's radiation
characteristics $\bF(\theta,\phi)$ by altering the antenna's physical configuration $\br'$.
Thus, it is possible to have orthogonal radiation characteristics, and the numbers of distinct radiation states at the transmitter and the receiver, $Q$ and $W$,  correspond to the orthogonal radiation characteristics that the reconfigurable antennas can generate, which are determined by the practical circuit and antenna designs.
We further assume that all the antenna ports can be reconfigured simultaneously. Then, the total number of possible combinations in which the transmit and receive ports can be reconfigured is given by $\Psi=QW$. We refer to each one of these combinations as a reconfiguration state.
For brevity, we further refer to the $\psi$-th reconfiguration state as reconfiguration state $\psi$.
Note that the transmitter and the receiver may have the same or different reconfiguration capabilities in practice, and our analysis is applicable to both cases by adjusting the number of radiation states at the transmitter $Q$ and the number of radiation states at the receiver $W$. When $Q=W$, the transmitter and the receiver have the same reconfiguration capability. Otherwise, the transmitter and the receiver have different reconfiguration capabilities. 

We consider the narrowband block-fading channels.
Denote the transmitted signal vector from the transmitter as $\bx\in \DC^{N_t\times 1}$ with a transmit power constraint $\Tr\left(\mathbb{E}\{\bx\bx^H\}\right)=P$. The received signal at the receiver with reconfiguration state $\psi$ is given by
\begin{equation}\label{eq:yhxbasic}
  \by=\bH_{\psi}\bx+\bn,
\end{equation}
where $\bH_{\psi}\in\DC^{N_r\times N_t}$ denotes the channel matrix corresponding to the reconfiguration state $\psi$ and $\bn\sim\mathcal{CN}(\mathbf{0};\sigma^2_n\bI_{N_r})$ denotes the additive white Gaussian noise (AWGN) vector at the receive antennas.
Without loss of generality, the noise variance is taken to be unity, i.e., $\sigma^2_n=1$.
Note that $\bH_{\psi}(i,j)$
represents the channel coefficient that contains
the gain and phase information of the path between the $i$-th
receive antenna and the $j$-th transmit antenna in the  reconfiguration state $\psi$.
We assume that the channel matrices for different reconfiguration states are independent~\cite{Grau_08_AreMIMOcym,Fazel09stsbcmmioyra,Vakilian_15_SThmmra,Vakilian_15_ThmmraAr}, and have the same average channel power such that $\mathbb{E}\{\left\|\bH_{1}\right\|^2_F\}=\cdots=\mathbb{E}\{\left\|\bH_{\Psi}\right\|^2_F\}=N_rN_t$.
It is worth mentioning that the assumption of orthogonal radiation characteristics is crucial for our analysis, since our results are based on the assumption of independent channel matrices associated with different reconfiguration states. If the radiation patterns are non-orthogonal, the channel matrices associated with different reconfiguration states would be correlated, and we would expect a relatively smaller throughput gain compared with the results in this paper.

We further assume that full channel state information (CSI)  of all reconfiguration states is known at the receiver.  The full CSI is not necessarily known at the transmitter. We assume that a limited feedback from the receiver to the transmitter is available for the  reconfiguration state selection and transmit beam selection, which will be detailed later in Section~\ref{Sec:Architecture}.
We would like to point out that the assumption of full CSI at the receiver or even at both the transmitter and the receiver has often been adopted in the existing papers on mmWave systems and reconfigurable antennas, see, e.g.,~\cite{Sohrabi7913599,Rusu7579557,Sohrabi7389996,Chen7055330,Amadori_15_LowRDBStion,Ayach_14_SpatiallySparsePrecodingmmMIMO,Grau_08_AreMIMOcym,Fazel09stsbcmmioyra,Vakilian_15_SThmmra,Vakilian_15_ThmmraAr}.
On the other hand, it is worth mentioning that the full CSI assumption is not easy to achieve in practice. The full CSI requires user equipments to conduct channel estimation. However, channel estimation is relatively challenging for mmWave MIMO systems. Different from sub-6 GHz systems, the precoder for mmWave MIMO with a limited number of RF chains is usually not fully digital.
Due to the constrained precoding structure, the channel has to be estimated via the use of a certain number of RF beams, and each beam only presents a projection of the channel matrix rather than the full channel matrix itself. To obtain the full CSI, the channel estimation needs to be conducted over enough number of RF beams so that the full channel matrix can be estimated. The same process has to be repeatedly conducted for all the reconfiguration states as well. Thus, the channel estimation process to obtain full CSI would incur a considerable latency, and the assumption of full CSI is not easy to achieve in practice.




\subsection{Channel Model}

For low-frequency $N_r\times N_t$ MIMO systems in an ideally rich scattering environment, the channel for each reconfiguration state is usually modelled by a full rank $N_r\times N_t$ matrix with i.i.d. entries~\cite{Grau_08_AreMIMOcym}, e.g., i.i.d. complex Gaussian entries for Rayleigh fading channels.
For mmWave communications in the clustered scattering environment, it is no longer appropriate to model the channel for each reconfiguration state as a full-rank matrix with i.i.d. entries due to the sparse nature of mmWave channels.\footnote{The existing studies on the 2$\times$2 mmWave MIMO with reconfigurable antennas used the idealized assumption of full-rank channel matrices with i.i.d. complex Gaussian entries for all reconfiguration states, which is not practical for the general mmWave scenarios where the number of antennas is relatively large~\cite{Vakilian_15_SThmmra,Vakilian_15_ThmmraAr}.}
In the following, we present the channel model of mmWave MIMO systems with reconfigurable antennas.

\subsubsection{Physical Channel Representation}
The physical channel representation is also often known as Saleh-Valenzuela (S-V) geometric model.
The mmWave MIMO channel can be characterized by physical multipath models. In particular,
the clustered channel representation is usually adopted as a practical model for mmWave channels~\cite{akdeniz2014millimeter,Gustafson_14_ommcacm,Health_16_OverviewSPTmmMIMO}.
The channel matrix for reconfiguration state $\psi$ is contributed by $N_{\psi,\cl}$ scattering clusters, and each cluster contains $N_{\psi,\ry}$ propagation paths.
The 2D physical multipath model for the channel matrix $\bH_{\psi}$ is given by
\begin{equation}\label{eq:H_PhyscialModeling2D}
  \mathbf{H_\psi}=
   \sum^{N_{\psi,\cl}}_{i=1}\sum^{N_{\psi,\ry}}_{l=1}
   \alpha_{\psi,i,l} \mathbf{a}_{R}\left(\theta^r_{\psi,i,l}\right)
  \mathbf{a}_{T}^H\left(\theta^t_{\psi,i,l}\right),
\end{equation}
where
$\alpha_{\psi,i,l}$ denotes the path gain, $\theta^r_{\psi,i,l}$ and $\theta^t_{\psi,i,l}$ denote the angle of arrival (AOA) and the angle of departure (AOD), respectively,
$\mathbf{a}_{R}\left(\theta^r_{\psi,i,l}\right)$ and $\mathbf{a}_{T}^H\left(\theta^t_{\psi,i,l}\right)$ denote the steering vectors of the receive antenna array and the transmit antenna array, respectively.
In this work, we consider the 1D uniform linear array (ULA) at both the transmitter and the receiver. Then, the steering vectors are given by
\begin{equation}\label{}
  \mathbf{a}_{R}\left(\theta^r_{\psi,i,l}\right)=\left[1,e^{-j2\pi\vartheta^r_{\psi,i,l}},\cdots,e^{-j2\pi\vartheta^r_{\psi,i,l}(N_r-1)}\right]^T,
\end{equation}
and
\begin{equation}\label{}
    \mathbf{a}_{T}\left(\theta^t_{\psi,i,l}\right)=\left[1,e^{-j2\pi\vartheta^t_{\psi,i,l}},\cdots,e^{-j2\pi\vartheta^t_{\psi,i,l}(N_t-1)}\right]^T,
\end{equation}
where $\vartheta$ denotes the normalized spatial angle.  The normalized spatial angle is related to the physical AOA or AOD $\theta\in\left[-\pi/2,\pi/2\right]$ by
$
  \vartheta={d\sin(\theta)}/{\lambda},
$
where $d$ denotes the antenna spacing and $\lambda$ denotes the wavelength.


We assume that $N_{1,\cl}=\cdots=N_{\Psi,\cl}$ and $N_{1,\ry}=\cdots=N_{\Psi,\ry}$, which  implies that the sparsity of the mmWave MIMO channel remains the same for all reconfiguration states. 
By transmitting and receiving with orthogonal radiation patterns, the propagated signals undergo different reflections and diffractions such that different reconfiguration states lead to different multipath parameters. That is,
the values of $\alpha_{\psi,i,l}$, $\theta^r_{\psi,i,l}$, and $\theta^t_{\psi,i,l}$ change as the reconfiguration state changes.


\subsubsection{Virtual Channel Representation (VCR)}
The virtual (beamspace) representation is a natural choice for modelling  mmWave MIMO channels due to the highly directional nature of propagation~\cite{Health_16_OverviewSPTmmMIMO}.
The virtual model characterizes the physical channel by coupling between the
spatial beams in fixed virtual transmit and receive directions, and represents the channel in beamspace domain.

The VCR of $\mathbf{H_\psi}$ in~\eqref{eq:H_PhyscialModeling2D} is given by~\cite{Sayeed_02_Deconstuctingmfc,Tse_05_Fundamentals}
\begin{align}\label{eq:H_VirtualModeling}
  \mathbf{H_{\psi}}&=\sum^{N_r}_{i=1}\sum^{N_t}_{j=1} H_{\psi,V}(i,j)\mathbf{a}_R\left(\ddot{\theta}_{R,i}\right)
  \mathbf{a}_T^H\left(\ddot{\theta}_{T,j}\right) \notag\\
  &=\bA_R\mathbf{H}_{\psi,V}\bA_T^H,
\end{align}
where $\ddot{\theta}_{R,i}=\arcsin\left(\lambda\ddot{\vartheta}_{R,i}/d\right)$ and $\ddot{\theta}_{T,j}=\arcsin\left(\lambda\ddot{\vartheta}_{T,j}/d\right)$
are fixed virtual receive and transmit angles corresponding to uniformly spaced spatial angles\footnote{Without loss of generality, we here assume that $N_r$ and $N_t$ are odd.}
\begin{equation}\label{}
  \ddot{\vartheta}_{R,i}=\frac{i-1-(N_r-1)/2}{N_r}
\end{equation}
and
\begin{equation}\label{}
  \ddot{\vartheta}_{T,j}=\frac{j-1-(N_t-1)/2}{N_t},
\end{equation}
respectively,
\begin{equation}\label{}
  \bA_R=\frac{1}{\sqrt{N_r}}\left[ \mathbf{a}_R\left(\ddot{\theta}_{R,1}\right),\cdots,\mathbf{a}_R\left(\ddot{\theta}_{R,N_r}\right)\right]^T
\end{equation}
and
\begin{equation}\label{}
  \bA_T=\frac{1}{\sqrt{N_t}}\left[ \mathbf{a}_T\left(\ddot{\theta}_{T,1}\right),\cdots,\mathbf{a}_T\left(\ddot{\theta}_{T,N_t}\right)\right]^T
\end{equation}
are unitary DFT matrices, and $\bH_{\psi,V}\in\DC^{N_r\times N_t}$ is the virtual channel matrix.
Since $\mathbf{A}_R\mathbf{A}_R^H=\mathbf{A}_R^H\mathbf{A}_R=\bI_{N_r}$ and $\mathbf{A}_T \mathbf{A}_T^H= \mathbf{A}_T^H \mathbf{A}_T=\bI_{N_t}$, the virtual channel matrix and the physical channel matrix are unitarily equivalent, such that
 \begin{equation}\label{}
  \HV=\mathbf{A}_R^H\mathbf{H}_{\psi}\mathbf{A}_T.
\end{equation}




\subsubsection{Low-Dimensional VCR}
For MIMO systems, the link capacity of the reconfiguration state $\psi$ is directly related to
the amount of scattering and reflection in the multipath environment.
As discussed in~\cite[Chapter 7.3]{Tse_05_Fundamentals}, the number of non-vanishing rows and columns of $\HV$ depends on the amount of scattering and reflection. 
In the clustered scattering environment of mmWave MIMO, the dominant channel power is expected to be captured by a few rows and columns of the virtual channel matrix, i.e., a low-dimensional submatrix of $\HV$.

The discussion above  motivates the development of low-dimensional virtual representation of mmWave MIMO channels and the corresponding low-complexity beamforming designs for mmWave MIMO transceivers~\cite{Brady_13_BeamspaceSAMAM,Amadori_15_LowRDBStion,Sayeed_07_maxMcsparseRAA,Raghavan_11_SublinearSparse}.
Specifically, a low-dimensional virtual channel matrix, denoted by $\sHV\in\DC^{L_r\times L_t}$,
is obtained by beam selection from $\HV$, such that $\sHV$ captures $L_t$ dominant transmit beams and $L_r$ dominant receive beams of the full virtual channel matrix.
The low-dimensional virtual channel matrix is defined by
\begin{equation}\label{eq:sHv1}
  \sHV=\left[\HV(i,j)\right]_{i\in{\mathcal{M}_{\psi,r}},j\in\mathcal{M}_{\psi,t}},
\end{equation}
where $\mathcal{M}_{\psi,r}=\left\{i:(i,j)\in\mathcal{M}_{\psi}\right\}$, $\mathcal{M}_{\psi,t}=\left\{j:(i,j)\in\mathcal{M}_{\psi}\right\}$, and  $\mathcal{M}_{\psi}$ is the beam selection mask.
The beam selection mask $\mathcal{M}$ is related to the criterion of beam selection. For example, a common beam selection is based on the criterion of maximum magnitude, and the corresponding beam selection mask is defined as~\cite{Brady_13_BeamspaceSAMAM}
\begin{equation}\label{eq:Mmagnit}
  \mathcal{M}_{\psi}=\left\{(i,j):\left|\HV(i,j)\right|^2\ge\gamma_{\psi}\max_{(i,j)}\left|\HV(i,j)\right|^2\right\},
\end{equation}
where $0<\gamma_{\psi}<1$ is a threshold parameter used to ensure that $\sHV$ has the dimension of $L_r\times L_t$. Using \eqref{eq:Mmagnit}, the resulting $\sHV$ captures a fraction $\gamma_{\psi}$ of the power of $\HV$.

Although the VCR and the beamspace hybrid beamforming for mmWave systems have been widely adopted and studied in the literature~\cite{Health_16_OverviewSPTmmMIMO,Gao8284058,Brady_13_BeamspaceSAMAM,Amadori_15_LowRDBStion,Wang7974749,Mo7094595}, it is worth mentioning that there are some potential limitations on the utility of  VCR in practical mmWave systems.
For sub-6 GHz systems, motivated by the limitations of the S-V model and the statistical model, the intermediate VCR was introduced to keep the essence of S-V model without its complexity and to provide a tractable channel characterization~\cite{Sayeed_02_Deconstuctingmfc,Raghavan4487419,Huang5074791}. The VCR for sub-6 GHz systems  offers a simple and transparent interpretation of the effects of scattering and array characteristics~\cite{Sayeed_02_Deconstuctingmfc}. However, the interpretation of VCR is based on the assumption of (approximately) unitary bases at both the transmit and the receive sides, which may not be possible  in many practical mmWave systems.
Note that, different from sub-6 GHz systems, mmWave systems usually do not have a large number of dominant clusters. Also, the numbers of antennas for practical mmWave transceivers are still far from the asymptotics.
Thus, there may not be a solid case for the VCR in practical mmWave systems, and the statistics of the VCR for sub-6 GHz systems  do not hold for mmWave systems.
Consequently, the optimality of the beamspace hybrid beamforming, which is based on the VCR, in practical mmWave systems may not be easy to verify. The losses related to the application of the  simple DFT and IDFT are not clear. In addition, the power leakage issue due to the beam selection may be serious in practical mmWave systems. While the DFT and IDFT are fixed, the actual angles of departure and arrival of different paths are continuously distributed. Thus, the power of a path can leak into multiple different beams~\cite{Brady_13_BeamspaceSAMAM}, and selecting only a few number of beams may result in a serious issue of power loss.

\subsection{Transceiver Architecture}\label{Sec:Architecture}
For wireless communications at low frequencies, signal processing happens in the baseband, and MIMO relies on digital beamforming.
At mmWave frequencies, it is difficult to employ a separate RF chain and data converter for each antenna, especially in large MIMO systems, due to the complicated hardware implementation, the high power consumption, and/or the prohibitive cost~\cite{Pi11Aninmmvmbs,Doan04dcf60gcmosra,Ayach_14_SpatiallySparsePrecodingmmMIMO}.
To reduce the  number of RF chains and the number of data converters,
mmWave MIMO transceivers usually have low-complexity architectures and signal processors, e.g., analog beamformer, hybrid analog-digital beamformer, and low resolution transceivers.
In this work, we adopt the reconfigurable beamspace hybrid beamformer as the architecture of low-complexity transceivers for mmWave MIMO systems with reconfigurable antennas. 




At the transmitter, the  symbol vector $\bs\in\DC^{N_s\times 1}$ is first processed by a low-dimensional digital precoder $\bF\in\DC^{ L_t\times N_s}$, where $L_t$ denotes the number of RF chains at the transmitter. The obtained $L_t\times 1$ signal vector is denoted by $\sxV=\bF\bs$, which is then converted to analog signals by $L_t$ digital-to-analog converters (DACs). Next, the $L_t$ signals go through the beam selector to obtain the $N_t\times 1$ (virtual) signal vector $\bx_V$.
For a given beam selection mask $\Mset$, $\bx_V$ is constructed by $\left[\bx_V(j)\right]_{j\in\MsetT}=\sxV$ and $\left[\bx_V(j)\right]_{j\notin\MsetT}=\mathbf{0}$, where $\MsetT=\left\{j:(i,j)\in\Mset\right\}$.
The beam selector can be easily realized by switches in practice. $\bx_V$ is further processed by the DFT analog precoder $\bA_T\in\DC^{N_t\times N_t}$, and the obtained signal vector is given by $\bx=\bA_T\bx_V$.
Note that $\Tr\left(\mathbb{E}\{\sxV\sxV^H\}\right)=\Tr\left(\mathbb{E}\{\bx_V\bx_V^H\}\right)=\Tr\left(\mathbb{E}\{\bx\bx^H\}\right)=P$.
Finally, the transmitter sends $\bx$ with the reconfigurable antennas. 

The received signal vector at the receive antennas with a given reconfiguration state $\psi$ is given by
\begin{equation}\label{}
  \by=\bH_{\psi}\bx+\bn=\bA_R\HV\bA_T^H\bx+\bn=\bA_R\HV\bx_V+\bn.
\end{equation}
At the receiver side, $\by$ is first processed by the IDFT analog decoder $\bA_R^H\in\DC^{N_r\times N_r}$, and the obtained (virtual) signal vector is given by
\begin{equation}\label{eq:virtualsystemrepre}
 \by_V=\bA_R^H\by=\HV\bx_V+\bn_V,
\end{equation}
where the distribution of $\bn_V=\bA_R^H\bn$ is $\mathcal{CN}(\mathbf{0};\sigma^2_n\bI_{N_r})$.
Note that the system representation in~\eqref{eq:virtualsystemrepre} is unitarily equivalent to the antenna domain representation in~\eqref{eq:yhxbasic}.

According to the given beam selection mask $\Mset$,  the receiver then uses the beam selector to obtain the low-dimensional $L_r\times 1$ signal vector $\syV=\left[\by_V(i)\right]_{i\in\MsetR}$, where $L_r$ denotes the number of the RF chains at the receiver and $\MsetR=\left\{i:(i,j)\in\Mset \right\}$.
The low-dimensional virtual system representation for a given reconfiguration state $\psi$ is formulated as
\begin{equation}\label{eq:lowvirsyseq}
  \syV=\sHV\sxV+\snV,
\end{equation}
where $\sHV=\left[\cHV(i,j)\right]_{i\in{\MsetR},j\in\MsetT}$, $\snV=\left[\bn_V(i)\right]_{i\in\MsetR}$,
and $\snV\sim\mathcal{CN}(\mathbf{0};\sigma^2_n\bI_{L_r})$.
The analog signals are finally converted to digital signals by $L_r$ analog-to-digital converters (ADCs) for the low-dimensional digital signal processing.

As mentioned earlier, we assume that the full CSI is perfectly known at the receiver, and a limited feedback is available from the receiver to the transmitter to enable the beam selection and the reconfiguration state selection. Per the number of all possible combinations of selected beams and reconfiguration states, the number of the feedback bits is equal to $\log_2\left(\Psi\right)+\log_2\left(\binom{N_t}{L_t}\binom{N_r}{L_r}\right)$.
We assume that $N_s=L_t\le L_r$ to maximize the multiplexing gain of the system.
The digital precoder at the transmitter is then given by $\bF=\bI_{N_s}$  with equal power allocation between the $N_s$ data streams, since the transmitter does not have the full CSI.
At the receiver, the digital decoder is the joint ML decoder for maximizing the throughput.\footnote{The relatively low-complexity MMSE-SIC decoder can also be adopted here to maximize the average throughput.}
With the aforementioned transceiver architecture and CSI assumptions, the system throughput with a selected $\sHV$ is given by~\cite{Tse_05_Fundamentals}
\begin{equation}\label{eq:Coptimal}
  R_{\sHV}=\log_2\left|\bI_{L_r}+\frac{\rho}{L_t}\sHV\sHV^H\right|,
\end{equation}
where $\rho=P/\sigma^2_n$ denotes the transmit power to noise ratio.

%

\section{Throughput Gain of Employing Reconfigurable Antennas}\label{sec:thrgainana}

In this section, we analyze the performance gain of employing the reconfigurable antennas in terms of the throughput.
With the optimal reconfiguration state selection, the instantaneous system throughput is given by
\begin{equation}\label{eq:defRsc}
R_{\cpsi}=\max_{\psi\in\left\{1,\cdots,\Psi\right\}}R_{\psi},
\end{equation}
where
\begin{align}\label{eq:defRs}
  R_{\psi}&=\log_2\left|\bI_{L_r}+\frac{\rho}{L_t}\cbHV\cbHV^H\right|\notag\\
  &=\max_{\sHV\in\left\{\tilde{\mathcal{H}}_\psi\right\}}\log_2\left|\bI_{L_r}+\frac{\rho}{L_t}\sHV\sHV^H\right|
\end{align}
represents the maximum achievable throughput under the reconfiguration state $\psi$,
$\cssHV$ denotes the optimal low-dimensional virtual channel of  $\HV$,
and $\tilde{\mathcal{H}}_\psi$ denotes the set of all possible $L_r\times L_t$ submatrices of $\HV$.
Here, the optimal reconfiguration state is the reconfiguration state that maximizes the throughput.

\subsection{Average Throughput Gain}\label{sec:appAvethrougai}

The average throughput gain of employing the reconfigurable antennas is given by
\begin{equation}\label{eq:th_gain}
 G_{\bar{R}}={\bar{R}_{\cpsi}}/{\bar{R}_{\psi}}, 
\end{equation}
where $\bar{R}_{\cpsi}=\mathbb{E}\{R_{\cpsi}\}$, $\bar{R}_{\psi}=\mathbb{E}\{R_{\psi}\}$, and the expectation is over different channel realizations.
As mentioned before, we assume that the channel matrices for different reconfiguration states have the same average channel power, and hence,  $\bar{R}_{1}=\cdots=\bar{R}_{\Psi}$.

With the key property of VCR, each entry of $\sHV$, i.e., $\sHV(i,j)$, is associated with a set of physical paths~\cite{Sayeed_02_Deconstuctingmfc}, and it is approximated equal to the sum of the complex gains of the corresponding paths~\cite{Sayeed_07_maxMcsparseRAA}. When the number of distinct paths associated with $\sHV(i,j)$ is sufficiently large, we note from the central limit theorem that $\sHV(i,j)$ tends toward a complex Gaussian random variable. As observed in~\cite{Gustafson_14_ommcacm} for the practical mmWave propagation environment at 60 GHz, the average number of distinct clusters is 10, and the average number of rays in each cluster is 9. The 802.11ad model has a fixed value of 18 clusters for the 60 GHz WLAN systems~\cite{Maltsev_10_cmf60gwsmodl}.
With the aforementioned numbers of clusters and rays, the entries of $\sHV$ can be approximated by zero-mean complex Gaussian variables.
Different from the rich scattering environment for low-frequency communication, the associated groups of paths to different entries of  $\sHV$ may be correlated in the mmWave environment. As a result, the entries of  $\sHV$ can be correlated, and the entries of $\sHV$ are then approximated by correlated  zero-mean complex Gaussian variables.
In the literature, it has been shown that the instantaneous capacity of a MIMO system whose channel matrix has correlated  zero-mean complex Gaussian entries can be approximated by a Gaussian variable~\cite{Moustakas_03_Mctccitpocian,Martin_03_aedacfcucf}. Based on the discussion above,
the distribution of $R_{\psi}$ is approximated by a Gaussian distribution, and the accuracy of 
the approximation
will also be numerically shown later in Section~\ref{sec:numersim}.
It is worth mentioning that practical mmWave channels may have relatively small numbers of clusters and paths. Different from the results in~\cite{Gustafson_14_ommcacm,Maltsev_10_cmf60gwsmodl} for 60 GHz, one may observe only 3-6 clusters at 28 GHz~\cite{Raghavan8255763,Raghavan8053813}. Although the distribution of $\sHV(i,j)$ may deviate from Gaussian when the number of distinct paths associated with it becomes small, we find that $R_{\psi}$ is still approximately Gaussian distributed. Note that Gaussian approximated distribution for the rate of MIMO systems has been shown many times in the literature under various assumptions~\cite{Telatar_99_CmGaucs,Moustakas_03_Mctccitpocian,Smith_04_Aappcdfms,Martin_03_aedacfcucf}, and the Gaussianity of $R_{\psi}$  with randomness in the angular profiles of the clusters is reasonable.

Denoting the approximated Gaussian distribution of $R_{\psi}$ as $\mathcal{N}(\muR,\varR)$,
where $\muR$ and $\varR$ denote the mean and the variance of $R_\psi$, respectively,
we have the following proposition giving the approximated average throughput gain of employing the reconfigurable antennas.

\begin{Proposition}\label{Prop:1}
The average throughput gain of employing the reconfigurable antennas with $\Psi$ distinct reconfiguration states is approximated by
\begin{equation}\label{eq:th_gain_close}
  G_{\bar{R}} \approx\int_0^\infty \frac{1}{\muR}-\frac{1}{2^\Psi\muR}\left(1+\mathrm{erf}\left(\frac{x-\muR}{\sqrt{2\varR}}\right)\right)^\Psi\mathrm{d}x. 
\end{equation}
\end{Proposition}
\begin{IEEEproof}
See Appendix~\ref{App:proofaveGa}
\end{IEEEproof}


To the best of our knowledge, the expression for $G_{\bar{R}}$ in~\eqref{eq:th_gain_close} cannot be simplified for general values of $\Psi$.
Thus, the calculation of the average throughput gain for a general number of reconfiguration states involves an infinite integral of a complicated function.  
We then further consider two special cases, in which the relatively simple expressions for the average throughput gain are tractable.


\subsubsection{$\Psi\le5$}
In practice, the number of distinct reconfiguration states is usually small, due to the complicated hardware design and the limited size of antennas. The following Corollary gives the approximated average throughput gain for the case of $\Psi\le5$.
\begin{Corollary}\label{Cor:aveGst5}
The approximated expressions for the average throughput gain,  $G_{\bar{R}}$, for the case of $\Psi\le5$ are given by
\begin{align}\label{eq:th_gain_close_125}
 &G_{\bar{R}}(\psi=1)\approx1,~
 G_{\bar{R}}(\psi=2)\approx1+\frac{1}{\muR}\sqrt{\frac{\varR}{\pi}},\notag\\
  &G_{\bar{R}}(\psi=3)\approx1+\frac{3}{2\muR}\sqrt{\frac{\varR}{\pi}},\notag\\
 &G_{\bar{R}}(\psi=4)\approx1+\frac{3}{\muR}\sqrt{\frac{\varR}{\pi^3}}\arccos\left(-\frac{1}{3}\right),\notag\\
 &G_{\bar{R}}(\psi=5)\approx1+\frac{5}{2\muR}\sqrt{\frac{\varR}{\pi^3}}\arccos\left(-\frac{23}{27}\right).
\end{align}

\end{Corollary}
\begin{IEEEproof}
See Appendix~\ref{App:proofaveGast5}
\end{IEEEproof}



\subsubsection{$\Psi$ Is Large}
Although $\Psi$ is relatively  small in practice, it is of theoretical importance to study the case of large $\Psi$ to capture the limiting performance gain of reconfigurable antennas.
We present the approximated average throughput gain for large $\Psi$ in the corollary below.

\begin{Corollary}\label{Cor:aveGlsn}
The approximated  average throughput gain,  $G_{\bar{R}}$, for the case of large $\Psi$ is given by
\begin{align}\label{eq:GavelargePsi}
  &G_{\bar{R}}\approx 1+\frac{\sqrt{2\varR}}{\muR}\notag\\ &\left((1-\beta)\mathrm{erf}^{-1}\left(1-\frac{2}{\Psi}\right)
  +\beta \mathrm{erf}^{-1}\left(1-\frac{2}{e\Psi}\right)\right),
\end{align}
where $\beta\simeq0.5772$ denotes the Euler's constant.
\end{Corollary}
\begin{IEEEproof}
See Appendix~\ref{App:proofaveGlsn}
\end{IEEEproof}

Based on~\eqref{eq:GavelargePsi}, we further obtain the limiting behavior of the average throughput gain when $\Psi$ becomes large in the following corollary.
 \begin{Corollary}\label{Cor:aveGlsngrO}
As $\Psi\rightarrow\infty$,  $G_{\bar{R}}(\Psi)$ is asymptotically equivalent to
\begin{equation}\label{eq:galarpsiasy}
G_{\bar{R}}(\Psi)\sim\frac{\sqrt{2\varR}}{\muR}\sqrt{\ln(\Psi)}.
\end{equation}
\end{Corollary}
\begin{IEEEproof}
See Appendix~\ref{App:proofCor:aveGlsngrO}
\end{IEEEproof}

From \eqref{eq:galarpsiasy} we find that $G_{\bar{R}}(\Psi)=O\left(\sqrt{\ln(\Psi)}\right)$ as $\Psi\rightarrow\infty$. Thus,  the growth of the average throughput from adding reconfiguration states  becomes small when the number of reconfiguration states  is already large, although having more distinct reconfiguration states always benefits the  average throughput.

\subsection{Outage Throughput Gain}
In the above analysis, we focused on the performance gain in terms of the average throughput.
However, it is insufficient to use the average throughput as the sole measure of the rate performance of the systems with multiple antennas. 
For scenarios where the channel remains  (quasi) static during the transmission, it is appropriate to evaluate the system performance by the outage throughput, since every possible target transmission rate is associated with an unavoidable probability of outage.
In the following, we analyze the performance gain of employing the reconfigurable antennas in terms of the outage throughput.

For a given target rate $R$, an outage event happens when the maximum achievable throughput is less than the target rate, and the outage probabilities for the systems without and with the reconfigurable antennas are given by
$\PP(R_{\cpsi}<R)$ and $\PP(R_{\psi}<R)$, respectively.
At a required outage level  $0<\epsilon<1$,  the outage throughputs for the systems  with and without the reconfigurable antennas are given by~\cite{Tse_05_Fundamentals}
\begin{align}\label{eq:outthsin}
 R_{\cpsi}^{\out}=\max~R,~~\mathrm{s.t.}~~\PP(R_{\cpsi}<R)\le\epsilon
\end{align}
and
\begin{align}\label{eq:outthrec}
 R_{\psi}^{\out}=\max~R,~~\mathrm{s.t.}~~\PP(R_{\psi}<R)\le\epsilon,
 \end{align}
respectively.
The outage throughput gain of employing the reconfigurable antennas at an outage level $\epsilon$ is given by
\begin{equation}\label{eq:Defth_gain_out}
 G_{R^\out}={R_{\cpsi}^{\out}}/{R_{\psi}^{\out}}, 
\end{equation}
where $R_{\cpsi}^{\out}$ and $R_{\psi}^{\out}$ are given in \eqref{eq:outthsin} and \eqref{eq:outthrec}, respectively.
The outage throughput gain of employing the reconfigurable antennas is given in the following proposition.
\begin{Proposition}\label{Prop:2}
The outage throughput gain of employing the reconfigurable antennas with $\Psi$ distinct reconfiguration states is approximated by
\begin{equation}\label{eq:th_gain_out}
 G_{R^\out}
\approx\frac{\muR-\sqrt{2\varR}\mathrm{erf}^{-1}\left(1-2\epsilon^{\frac{1}{\Psi}}\right)}{\muR-\sqrt{2\varR}\mathrm{erf}^{-1}\left(1-2\epsilon\right)}.
\end{equation}
\end{Proposition}
\begin{IEEEproof}
See Appendix~\ref{App:proofoutGa}
\end{IEEEproof}

\subsubsection{$\Psi$ Is Large}
We now investigate the limiting performance gain of reconfigurable antennas in terms of the outage throughput as $\Psi\rightarrow\infty$. 

Based on~\eqref{eq:th_gain_out}, we present the limiting behavior of the outage throughput gain when $\Psi$ becomes large in the following corollary.
 \begin{Corollary}\label{Cor:outGlsngrO}
 As $\Psi\rightarrow\infty$,  $G_{R^\out}(\Psi)$ is asymptotically equivalent to
\begin{equation}\label{eq:goutlarpsiasy}
  G_{R^\out}(\Psi)\sim\frac{\sqrt{2\varR}}{\muR-\sqrt{2\varR}\mathrm{erf}^{-1}\left(1-2\epsilon\right)}\sqrt{\ln(\Psi)}.
\end{equation}
\end{Corollary}
\begin{IEEEproof}
See Appendix~\ref{App:proofCor:outGlsngrO}
\end{IEEEproof}

Similar to the finding for the average throughput gain, we find from~\eqref{eq:goutlarpsiasy} that
$G_{R^\out}(\Psi)=O\left(\sqrt{\ln(\Psi)}\right)$ as $\Psi\rightarrow\infty$.
Thus, the growth of the outage throughput from adding reconfiguration states  becomes small when the number of reconfiguration states  is already large. Comparing the limiting behaviors of the average throughput gain and the outage throughput gain, we further find that the growth rate of the average throughput and the growth rate of the outage throughput have the same order when the number of reconfiguration states is large.

\section{Fast Selection Algorithm}\label{sec:fastalgrb}
In the previous section, we analyzed the throughput gain of employing the reconfigurable antennas when the optimal reconfiguration state and beams are selected, while the problem of how to select the optimal reconfiguration state and beams has not been considered. In fact, as will be discussed later, selecting the optimal  reconfiguration state and beams  among all possible selections is extremely complicated and challenging for practical applications. To overcome the challenge, in this section, we propose a fast selection algorithm with low complexity and near-optimal throughput performance in the sparse mmWave MIMO environment.

The objective of selecting the optimal reconfiguration state and beams is to obtain the corresponding optimal $\sHV$ that maximizes the system throughput given in~\eqref{eq:Coptimal}.
The design problem of selecting $\sHV$
is formulated as\footnote{Note that the selection criteria in~\eqref{eq:problemselect} is maximizing the throughput rather than maximizing the channel magnitude.} 
\begin{equation}\label{eq:problemselect}
  \max_{\psi\in\left\{1,\cdots,\Psi\right\}}\max_{\sHV\in\left\{\tilde{\mathcal{H}}_\psi\right\}}\left|\bI_{L_r}+\frac{\rho}{L_t}\sHV\sHV^H\right|.
\end{equation}

A straightforward method to obtain the optimal $\sHV$ is the exhaustive search among all possible selections of $\sHV$. That is, we first search for the optimal beam selection for each reconfiguration state to obtain $\cssHV$, i.e., the optimal low-dimensional virtual channel of $\HV$.
Then, we compare the obtained $\cssHV$ among all reconfiguration states to complete the selection of optimal $\sHV$, denoted by $\csHV$.
Since there are $\Psi$ reconfiguration states and $\binom{N_t}{L_t}\binom{N_r}{L_r}$ possible submatrices for each state, the total number of possible selections to search is given by
\begin{equation}\label{}
N_{\mathrm{total}}=\Psi\binom{N_t}{L_t}\binom{N_r}{L_r}=\frac{\Psi N_r!N_t!}{L_r!L_t!\left(N_r-L_r\right)!\left(N_t-L_t\right)!}.
\end{equation}
When $N_t\gg L_t$, $N_r\gg L_r$, and/or $\Psi\gg 1$, the total number to search, $N_{\mathrm{total}}$, would be too large for practical applications due to the high complexity.
Thus, in what follows, we propose a low-complexity design to obtain  $\csHV$ which achieves the near optimal throughput performance.


As discussed earlier, the mmWave MIMO channel has a sparse nature, and the number of non-vanishing rows and columns of the virtual channel matrix is relatively small in the clustered
scattering environment.
Now let us consider an extreme scenario such that all of the non-vanishing entries of $\HV$ are contained in the low-dimensional submatrix, and $\HV$ is approximated by
\begin{equation}\label{eq:ssHv}
 \bM\odot\HV,
\end{equation}
where
\begin{equation}\label{eq:MaskM}
  \bM(i,j)=\left\{
  \begin{array}{ll} 1\;, &\mbox{if}~(i,j)\in \widehat{\Mset}_{\psi},\\
  0\;, &\mbox{otherwise,}
  \end{array}
\right.
\end{equation}
and $\widehat{\Mset}_{\psi}$ is the beam selection mask corresponding to $\cssHV$. Note that a similar approximation was adopted in, e.g.,~\cite{Sayeed_07_maxMcsparseRAA} to approximate the sparse virtual MIMO channel. With~\eqref{eq:ssHv}, we have
\begin{align}\label{eq:appHtoHvl}
  &\left|\bI_{L_r}+\frac{\rho}{L_t} \cssHV \cssHV^H\right|
  \approx \left|\bI_{N_r}+\frac{\rho}{L_t} \HV\HV^H\right|\notag\\
  &=  \left|\bI_{N_r}+\frac{\rho}{L_t} \bH_{\psi}\bH_{\psi}^H\right|.
\end{align}
Based on~\eqref{eq:appHtoHvl}, we find that a fast selection of reconfiguration state can be achieved by directly comparing their (full) physical channel matrices. Instead of finding the optimal beam selection of each reconfiguration state first, we can directly determine the optimal reconfiguration state by \begin{equation}\label{eq:problemselect_2}
  \widehat{\psi}=\arg\max_{\psi\in\left\{1,\cdots,\Psi\right\}} \left|\bI_{N_r}+\frac{\rho}{L_t} \bH_{\psi}\bH_{\psi}^H\right|.
\end{equation}
Based on~\eqref{eq:problemselect_2}, the near-optimal reconfiguration state can be selected by calculating and comparing the throughput among only $\Psi$ possible channel matrices.
Note that $\bH_{\psi}$ in \eqref{eq:problemselect_2} is the full channel matrix rather than a low-dimensional virtual channel matrix associated with a particular selection of beams.
In addition, with the fast reconfiguration state selection, we can select beams from only the beams that are associated with the selected reconfiguration state. In contrast, the exhaustive search needs to examine the performance of all beams associated with all reconfigurable states.
Although the performance of this fast-selection method depends on the accuracy of the approximation in~\eqref{eq:ssHv}, we will show later  by numerical results that usually near optimal performance can be achieved.

\begin{algorithm}[!b]
\caption{Fast Selection}
\begin{algorithmic}[1]\label{Alg:1}
\begin{small}
\Procedure{FastSelAlg}{$\rho, N_r, N_t, L_r, L_t, \bH_{1}, \cdots, \bH_{\Psi}$}

\State
$\widehat{\psi}:=\arg\max_{\psi\in\left\{1,\cdots,\Psi\right\}} \left|\bI_{N_r}+\frac{\rho}{L_t} \bH_{\psi}\bH_{\psi}^H\right|$;

\State
$\mathcal{I}_r:=\left\{1, \cdots, N_r\right\}$;
$\mathcal{I}_t:=\left\{1, \cdots, N_t\right\}$;
$\bh_j:=j$-th row of $\cHV$, $\forall j\in\mathcal{I}_r$;

\State
$J:=\arg\max_{j\in\mathcal{I}_r}\bh_j\bh_j^H$;

\State
$\MsetR:=\left\{J\right\}$;
$\ssHV:=\left[\cHV\left(i,j\right)\right]_{i\in{\MsetR,j\in\mathcal{I}_t}}$;
$\mathcal{I}_r:=\mathcal{I}_r-\left\{J\right\};$

\For{$l:=2:L_r$}
\State
$J:=\arg\max_{j\in \mathcal{I}_r}\bh_j\left(\bI_{N_t}+\frac{\rho}{N_t}\ssHV^H\ssHV\right)^{-1}\bh_j^H$;

\State
$\MsetR:=\MsetR+\left\{J\right\}$;
$\ssHV:=\left[\cHV\left(i,j\right)\right]_{i\in{\MsetR,j\in\mathcal{I}_t}}$;
$\mathcal{I}_r:=\mathcal{I}_r-\left\{J\right\};$

\EndFor
\State
$\bh_j:=j$-th column of $\ssHV$, $\forall j\in\mathcal{I}_t$;

\State
$J:=\arg\max_{j\in\mathcal{I}_r}\bh_j^H\bh_j$;

\State
$\MsetT:=\left\{J\right\}$;
$\csHV:=\left[\ssHV\left(i,j\right)\right]_{i\in{\MsetR,j\in\MsetT}}$;
$\mathcal{I}_t:=\mathcal{I}_t-\left\{J\right\};$

\For{$l:=2:L_t$}
\State \vspace{-5.5mm}
\begin{align*}\label{}
 & J:=\arg\max_{j\in \mathcal{I}_t}\notag\\
&\bh_j^H\left(\bI_{L_r}-\frac{\rho}{L_t}\csHV\left(\bI_{l-1}+\frac{\rho}{L_t}\csHV^H\csHV\right)^{-1}\csHV^H\right)\bh_j;
\end{align*}

\State
$\MsetT:=\MsetT+\left\{J\right\}$;
$\csHV:=\left[\ssHV\left(i,j\right)\right]_{i\in{\MsetR,j\in\MsetT}}$;
$\mathcal{I}_t:=\mathcal{I}_t-\left\{J\right\};$

\EndFor
\State
\Return{ $\widehat{\psi}$, $\mathcal{M}_r$, $\mathcal{M}_t$, $\csHV$;}
\EndProcedure
\end{small}
\end{algorithmic}
\end{algorithm}

To reduce the complexity of beam selection, we utilize the technique of separate transmit and receive antenna selection~\cite{Sanayei_04_CMAHTRAS} and the incremental
successive selection algorithm (ISSA)~\cite{Alkhansari_04_fastassims}.
Again due to the sparsity of  mmWave channels, the dominant beams usually significantly outperform the other beams, and they can be easily selected by the fast beam selection scheme.
Note that the transmitter does not have the full CSI in the considered system, and hence, the existing beamspace selection schemes in, e.g., \cite{Amadori_15_LowRDBStion}, with the requirement of full CSI on the beamspace channel at the transmitter are not applicable in our work.
Our fast beam selection method is explained next.
The beam selection problem in fact includes both transmit and receive beam selections. We adopt a separate transmit and receive beam selection technique~\cite{Sanayei_04_CMAHTRAS} for first selecting the best $L_r$ receive beams and then selecting the best $L_t$ transmit beams. For both the receive and transmit beam selections, a technique based on  the incremental successive selection algorithm (ISSA)~\cite{Alkhansari_04_fastassims} is utilized.
We start from the empty set of selected beams, and then successively add an individual beam to this set at each step of the beam selection algorithm until the number of selected transmit and receive beams reach $L_t$ and $L_r$, respectively. Note that successively selecting one beam at each step for the algorithm does not mean that we successively turn on individual beams.
Instead, we first determine the set of selected beams by successively  selecting beams according to the algorithm. After determining the set of selected beams, the transmitter and the receiver selectively turn on all selected beams together by beam selectors~\cite{Brady_13_BeamspaceSAMAM,Health_16_OverviewSPTmmMIMO}.
In each step of the algorithm, the objective is to select one of the unselected beams that leads to the highest increase of the throughput.
The mechanism of ISSA-based receive beam selection is provided as follows.
Denote the submatrix corresponding to the selected $n$ receive beams after $n$ steps of ISSA as $\sHV^{n}\in\DC^{n\times N_t}$ and the $j$-th row of $\HV$ by $\bh_{\psi,V}^{j}$.
Since the contribution of the $j$-th receive beam to the throughput under the $\log$ 
function  is given by
\begin{equation}\label{eq:alphapsinjR}
  g_{\psi,n,j}=\bh_{\psi,V}^{j}\left(\bI_{N_t}+\frac{\rho}{N_t}\left(\sHV^{n}\right)^H\sHV^{n,r}\right)^{-1}\left(\bh_{\psi,V}^{j}\right)^H,
\end{equation}
we select the receive beam at the $n+1$ step by
\begin{equation}\label{}
 \bh_{\psi,V}^{J}=\arg\max_{\bh_{\psi,V}^{j}}  g_{\psi,n,j}.
\end{equation}
The mechanism of ISSA-based transmit beam selection is omitted here, since it is similar to that of the ISSA-based receive beam selection. 

The proposed fast selection algorithm is given as
Algorithm~1.  The outputs of the algorithm are the optimal reconfiguration state, the indices of the selected receive beams, the indices of the selected transmit beams, and the selected low-dimensional virtual channel, denoted by  $\widehat{\psi}$, $\mathcal{M}_r$, $\mathcal{M}_t$, and $\csHV$, respectively.

We would like to highlight that the proposed algorithm significantly reduces the complexity of reconfiguration state selection and beam selection, and achieves the near-optimal throughput performance.
It is worth pointing out that an important requirement of the proposed algorithm is the knowledge of full CSI of all reconfigurable states, and the associated channel estimation complexity has not been taken into account. Although the full CSI assumption has been widely-adopted in the literature, as mentioned earlier, the channel estimation is relatively challenging for mmWave systems with reconfigurable antennas.

\section{Numerical Results}\label{sec:numersim}
For all numerical results in this work, we adopt the clustered multipath channel model in~\eqref{eq:H_PhyscialModeling2D} to generate the channel matrix. We assume that $\alpha_{\psi,i,l}$ are i.i.d. $\mathcal{CN}\left(0,\sigma^2_{\alpha,\psi,i}\right)$, where $\sigma^2_{\alpha,\psi,i}$ denotes the average power of the $i$-th cluster,
and
$\sum_{i=1}^{N_{\psi,c}}\sigma^2_{\alpha,\psi,i}=\gamma_{\psi}$,  where $\gamma_{\psi}$ is a normalization parameter to ensure that $\mathbb{E}\{\left\|\bH_{\psi}\right\|^2_F\}=N_rN_t$.
We also assume that $\theta^r_{\psi,i,l}$ are uniformly distributed with
mean $\theta_{\psi,i}^r$ and a constant angular spread (standard deviation) $\sigma_{\theta^r}$. $\theta^t_{\psi,i,l}$ are uniformly distributed with
mean $\theta_{\psi,i}^t$ and a constant angular spread (standard deviation) $\sigma_{\theta^t}$. We further assume that $\theta_{\psi,i}^r$ and $\theta_{\psi,i}^t$ are both uniformly distributed within the range of $[-\pi/2, \pi/2]$.
Unless otherwise stated, the system parameters are $N_r=N_t=17, L_r=L_t=5, N_{\psi,\cl}=10, N_{\psi,\ry}=8,$ $\sigma_{\theta^r}=\sigma_{\theta^t}=3^\circ$, and $d/\lambda=1/2$.
All average results are  over 5,000 randomly generated channel realizations.
Note that $N_{\psi,\cl}=10$ and $N_{\psi,\ry}=8$ are based on the existing observations at 60 GHz in the literature~\cite{Gustafson_14_ommcacm,Maltsev_10_cmf60gwsmodl}, and practical mmWave channels at 28 GHz may have relatively small numbers of clusters and paths~\cite{Raghavan8255763,Raghavan8053813}.

\begin{figure}[!htbp]
    \centering
    \begin{subfigure}[t]{\columnwidth}
        \centering
        \includegraphics[width=.9\columnwidth]{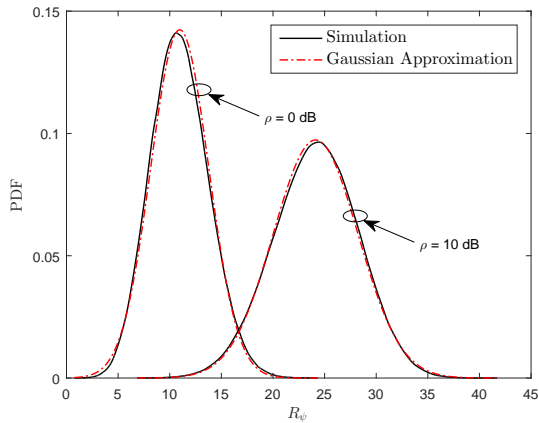}
        \caption{$N_{\psi,\cl}=10, N_{\psi,\ry}=8$.}
    \end{subfigure}\\
    \begin{subfigure}[t]{\columnwidth}
        \centering
        \includegraphics[width=.9\columnwidth]{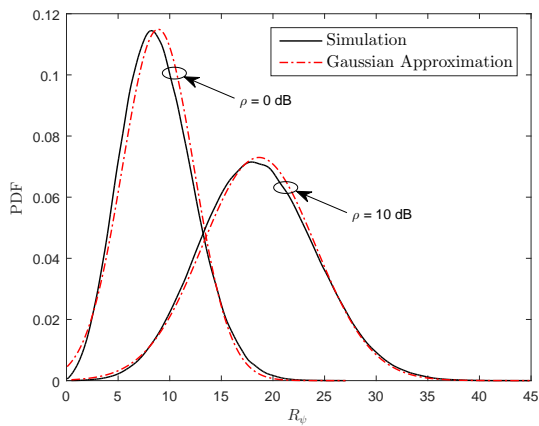}
        \caption{$N_{\psi,\cl}=4, N_{\psi,\ry}=2$.}
    \end{subfigure}
    \caption{PDF of $R_\psi$. The parameters are $N_r=N_t=17, L_r=L_t=5,  \sigma_{\theta^r}=\sigma_{\theta^t}=3^\circ$, and $d/\lambda=1/2$.}
    \label{fig:DisR}
\end{figure}


We first demonstrate the accuracy of the Gaussian approximated probability density function (PDF) of $R_\psi$.
Figure~\ref{fig:DisR} plots the simulated PDF and the Gaussian approximated PDF of $R_\psi$.
Figure~\ref{fig:DisR}(a) and  Figure~\ref{fig:DisR}(b) are for the case of $N_{\psi,\cl}=10$ and $N_{\psi,\ry}=8$ and the case of $N_{\psi,\cl}=4$ and $N_{\psi,\ry}=2$, respectively.
Different transmit power to noise ratios are considered, i.e., $\rho=0$~dB and $\rho=10$~dB.
The transmit powers of $30$~dBm and $40$~dBm are considered based on the  existing studies on mmWave systems \cite{Khan5876482,Pi11Aninmmvmbs,akdeniz2014millimeter}.
We note that the Gaussian approximations match the simulated PDFs.
In particular, we observe from Figure~\ref{fig:DisR}(b) that the distribution of $R_\psi$ can be well approximated by the Gaussian distribution even for relatively small numbers of clusters and paths.

\begin{figure}[!htbp]
\centering
\includegraphics[width=.9\columnwidth]{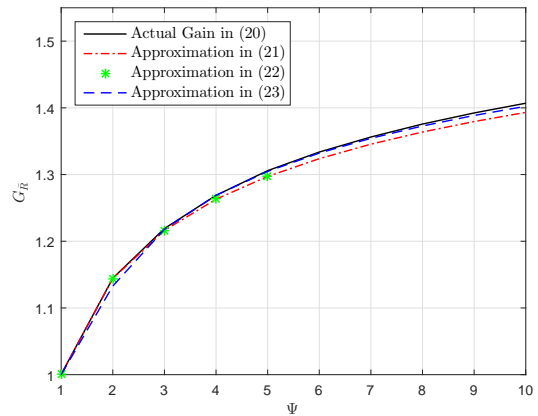}
\caption{Average throughput gain versus number of reconfiguration states. The parameters are $\rho=0$ dB, $N_r=N_t=17, L_r=L_t=5, N_{\psi,\cl}=10, N_{\psi,\ry}=8,$ $\sigma_{\theta^r}=\sigma_{\theta^t}=3^\circ$, and $d/\lambda=1/2$.}
\label{fig:Gain}
\end{figure}

We then show the average throughput gain of employing the reconfigurable antennas. Figure~\ref{fig:Gain} plots the average throughput gain, $G_{\bar{R}}$, versus the number of reconfiguration states, $\Psi$. The illustrated results are for the actual gain in~\eqref{eq:th_gain} by simulating the channels, $\bH_{\psi}$,
the theoretical approximation in~\eqref{eq:th_gain_close},
the simplified theoretical approximation for $\Psi\le5$ in~\eqref{eq:th_gain_close_125}, and the simplified theoretical approximation for large $\Psi$ in~\eqref{eq:GavelargePsi}. As depicted in the figure, the derived theoretical approximations match precisely the simulated results. 
In particular, we note that the simplified approximation for large $\Psi$ in~\eqref{eq:GavelargePsi} has good accuracy even when $\Psi$ is small.
From all four curves, we find that the growth of $G_{\bar{R}}$ with $\Psi$ is fast when $\Psi$ is small, while it becomes slow when $\Psi$ is relatively large.
This finding is consistent with the analysis in Section~\ref{sec:appAvethrougai}, and it indicates that the dominant average throughput gain of employing the reconfigurable antennas can be achieved by having a few number of reconfiguration states.

\begin{figure}[!htbp]
\centering
\includegraphics[width=.9\columnwidth]{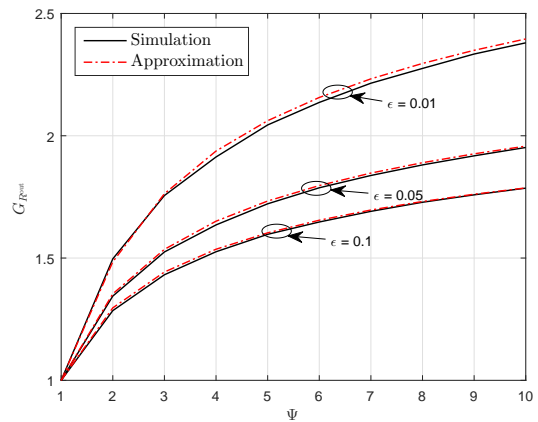}
\caption{Outage throughput gain versus number of reconfiguration states. The parameters are $\rho=0$ dB, $N_r=N_t=17, L_r=L_t=5, N_{\psi,\cl}=10, N_{\psi,\ry}=8,$ $\sigma_{\theta^r}=\sigma_{\theta^t}=3^\circ$, and $d/\lambda=1/2$.}
  \label{fig:GainOut}
\end{figure}


We now present the outage throughput gain of employing the reconfigurable antennas. Figure~\ref{fig:GainOut} plots the outage throughput gain, $G_{R^\out}$, versus the number of reconfiguration states, $\Psi$. Different outage levels are considered, i.e., $\epsilon=0.01, \epsilon=0.05$, and $\epsilon=0.1$. 
As the figure shows, $G_{R^\out}$ increases as $\Psi$ increases.
Similar to the results in Figure~\ref{fig:Gain}, we find that the dominant outage throughput gain of employing the reconfigurable antennas can be achieved by having a few number of reconfiguration states.
To obtain the outage throughput gain of $G_{R^\out}=1.5$, we only need $\Psi=2$, $\Psi=3$, and $\Psi=4$ reconfiguration states for the systems requiring $\epsilon=0.01, \epsilon=0.05$, and $\epsilon=0.1$, respectively. In addition, we note that $G_{R^\out}$ increases as $\epsilon$ increases, which indicates that the outage throughput gain of employing the reconfigurable antennas is more significant when the required outage level becomes more stringent.

\begin{figure}[!htbp]
\centering
\includegraphics[width=.9\columnwidth]{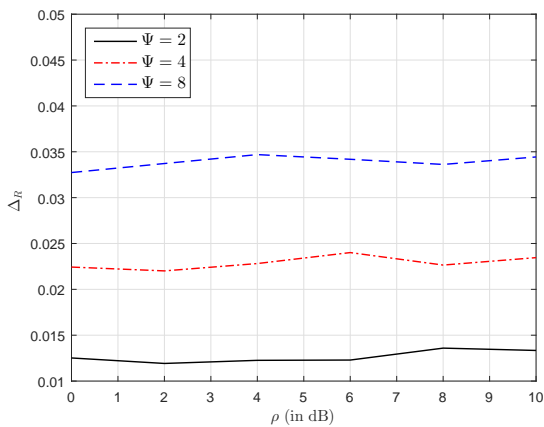}
\caption{Average throughput loss ratio versus transmit power to noise ratio. The parameters are $N_r=N_t=17, L_r=L_t=5, N_{\psi,\cl}=10, N_{\psi,\ry}=8,$ $\sigma_{\theta^r}=\sigma_{\theta^t}=3^\circ$, and $d/\lambda=1/2$. }
\vspace{-5mm}  \label{fig:Algerror}
\end{figure}

Finally, we examine the performance of the proposed algorithm for fast selection by evaluating the average throughput loss ratio, which is defined by
\begin{equation}\label{}
  \Delta_R=\left(\bar{R}_{\mathrm{max}}-\bar{R}_{\mathrm{fast}}\right)/{\bar{R}_{\mathrm{opt}}},
\end{equation}
where $\bar{R}_{\mathrm{max}}$ denotes the average throughput achieved by the exhaustive search and $\bar{R}_{\mathrm{fast}}$ denotes the average throughput achieved by the proposed fast selection algorithm.
Figure~\ref{fig:Algerror} plots the throughput loss ratio, $\Delta_R$, versus the transmit power to noise ratio, $\rho$. Systems with different numbers of reconfiguration states are considered, i.e., $\Psi=2$, $\Psi=4$, and  $\Psi=8$. As shown in the figure, the proposed fast selection algorithm achieves reasonably good performance compared with the maximum achievable throughput by the exhaustive search. Although $\Delta_R$ increases as $\Psi$ increases, the throughput loss ratio is less than $3.5\%$ even when $\Psi=8$.

\section{Conclusions and Future Work}\label{sec:concls}
In this paper, we have presented a framework for the theoretical study of the mmWave MIMO with reconfigurable antennas, where the low-complexity transceivers and the sparse channels are considered.
We have shown that employing reconfigurable antennas can provide both the average throughput gain and the outage throughput gain for mmWave MIMO systems. Also, we have derived the approximated expressions for the gains. Based on the highly sparse nature of mmWave channels, we have further developed a fast algorithm for reconfiguration state selection and beam selection. The accuracy of our derived expressions and the performance of the developed algorithm have been verified by numerical results. We have noted from the results that the dominant throughput gains by employing the reconfigurable antennas can be achieved by having a few number of reconfiguration states.
Numerical results have shown that a system with 3 reconfiguration states can achieve an
average throughput gain of 1.2 and an outage throughput gain of 1.5 for an outage requirement of $\epsilon=0.05$.

As a first comprehensive study on the mmWave MIMO with reconfigurable antennas, our paper can lead to a number of future research directions in this area.
In this paper, we have only considered the distinct reconfiguration states with independent channel matrices. In practice, the reconfiguration states may be dependent due to the non-orthogonal radiation characteristics, and the associated channel matrices may be correlated. Thus, taking the dependent reconfiguration states into consideration is an interesting future work, where we can have correlated and non-identically distributed $R_{\psi}$.
While this paper has adopted the VCR as the analytical channel model, as mentioned earlier, there are some potential limitations on the utility of  VCR in practical mmWave systems. Thus,  investigating the reconfigurable antennas for mmWave systems with other channel models, e.g., the S-V model, is an important and interesting future work.
Furthermore, future work on effective channel estimation techniques with low latency for mmWave systems with reconfigurable antennas is needed.
Also, the associated complexity of channel estimation needs to be taken into account for  future system designs.
For example, if the channel matrices associated with different reconfiguration states are (strongly) correlated, a potential scheme for reconfiguration state selection and beam selection  requiring a low-complexity channel estimation may start by setting the reconfiguration state to a random initial state. After finding the optimal beams in the initial state, one can reselect
another state based on the optimal beams for the initial state.

\appendices
\section{Proof of Proposition~\ref{Prop:1}}\label{App:proofaveGa}
With  the Gaussian approximation of the distribution of $R_{\psi}$, we can then approximate $R_{\cpsi}$  as the maximum of $\Psi$ i.i.d. Gaussian random variables, and
\begin{align}\label{eq:Rcpsibar}
\bar{R}_{\cpsi}&\approx\int_0^\infty 1-\left(F_{R_\psi}(x)\right)^\Psi \mathrm{d}x\notag\\
&=
\int_0^\infty 1-\frac{1}{2^\Psi}\left(1+\mathrm{erf}\left(\frac{x-\muR}{\sqrt{2\varR}}\right)\right)^\Psi\mathrm{d}x,
\end{align}
where
\begin{equation}\label{eq:CDFRpsi}
  F_{R_\psi}(x)=1+\mathrm{erf}\left(\frac{x-\muR}{\sqrt{2\varR}}\right)
\end{equation}
denotes the approximated cdf of $R_\psi$.

Substituting \eqref{eq:Rcpsibar} into \eqref{eq:th_gain} completes the proof.

\section{Proof of Corollary~\ref{Cor:aveGst5}}\label{App:proofaveGast5}
We rewrite $G_{\bar{R}}$ as
\begin{equation}\label{eq:th_gain_close_alt}
  G_{\bar{R}}(\Psi=i) \approx 1+\frac{\sqrt{\varR}}{\muR}E_{i}, 
\end{equation}
where $E_{i}$ denotes the mean of the maximum of $i$ independent standard normal random variables.
Denoting the cdf of standard normal distribution by $\Phi(\cdot)$, we have
\begin{align}\label{eq:proofeiniaj}
E_i&=\int_{-\infty}^\infty x\frac{\mathrm{d}\left(\Phi(x)\right)^i}{\mathrm{d}x}\mathrm{d}x \notag\\
&=i\left(i-1\right)\int_{-\infty}^\infty \frac{\exp\left(-x^2\right)}{2\pi}\left(\Phi(x)\right)^{i-2}\mathrm{d}x
\notag\\
&=\frac{i\left(i-1\right)}{2\pi}\sum_{j=0}^{\lfloor \frac{i}{2}-1\rfloor}\left(\frac{1}{2}\right)^{i-2-2j}\binom{i-2}{2j}A_j,
\end{align}
where $A_j=\int_{-\infty}^\infty \exp\left(-x^2\right)\left(\Phi(x)-\frac{1}{2}\right)^{2j}\mathrm{d}x $.
We note that $A_0$ and $A_1$ can be derived\footnote{The detailed steps to derive $A_0$ and $A_1$ are omitted here. The interested reader can find a relevant discussion at https://math.stackexchange.com/questions/473229.}, which are given by
$
  A_0=\sqrt{\pi}
$
and
$
  A_1=\frac{1}{2\sqrt{\pi}\tan^{-1}\left(\frac{\sqrt{2}}{4}\right)},
$
respectively.
Substituting $A_0$ and $A_1$ into \eqref{eq:proofeiniaj}, we can obtain $E_1=0$, $E_2=\pi^{-\frac{1}{2}}$, $E_3=\frac{3}{2}\pi^{-\frac{1}{2}}$, $E_4=3\pi^{-\frac{3}{2}}\arccos\left(-\frac{1}{3}\right)$,
and $E_5=\frac{5}{2}\pi^{-\frac{3}{2}}\arccos\left(-\frac{23}{27}\right)$. Finally, substituting the expressions for $E_1, \cdots, E_5$ into \eqref{eq:th_gain_close_alt} completes the proof.


\section{Proof of Corollary~\ref{Cor:aveGlsn}}\label{App:proofaveGlsn}
When $\Psi$ is large, we can adopt the Fisher-–Tippett theorem to approximate the distribution of the maximum of $\Psi$ independent standard normal random variables
as a Gumbel distribution, 
whose cumulative distribution
function (cdf) is given by
\begin{equation}\label{}
  F_{E_\Psi}(x)=\exp\left(-\exp\left(-\frac{x-\Phi^{-1}\left(1-\frac{1}{\Psi}\right)}{\Phi^{-1}\left(1-\frac{1}{e\Psi}\right)-\Phi^{-1}\left(1-\frac{1}{\Psi}\right)}\right)\right),
\end{equation}
where $\Phi^{-1}(\cdot)$ denotes the inverse cdf of the standard normal distribution.
We then have
\begin{align}\label{eq:Epsilarge}
E_\Psi &\approx 
\sqrt{2}\left(\left(1-\beta\right)\erf^{-1}\left(1-\frac{2}{\Psi}\right)+\beta\mathrm{erf}^{-1}\left(1-\frac{2}{e\Psi}\right)\right),
\end{align}
where $\beta$ denotes the Euler's constant.
Substituting \eqref{eq:Epsilarge} into \eqref{eq:th_gain_close_alt} completes the proof.

\section{Proof of Corollary~\ref{Cor:aveGlsngrO}}\label{App:proofCor:aveGlsngrO}
From the tail region approximation for the inverse error function, 
we note that
\begin{equation}\label{eq:appinverflargex}
  \mathrm{erf}^{-1}(x)=  \sqrt{-\ln\left(1-x^2\right)}  \quad \text{as} \quad x\rightarrow 1.
\end{equation}
Based on \eqref{eq:appinverflargex} and~\eqref{eq:GavelargePsi}, as $\Psi\rightarrow\infty$,  $G_{\bar{R}}(\Psi)$ is asymptotically equivalent to
\begin{align}
G_{\bar{R}}(\Psi)&\sim 1+\frac{\sqrt{2\varR}}{\muR}\left( \left(1-\beta\right)\sqrt{-\ln(4)+\ln\left(\frac{\Psi^2}{\Psi-1}\right)}\right.\notag\\
&\left.
+\beta\sqrt{1-\ln(4)+\ln\left(\frac{\Psi^2}{\Psi-1/e}\right)}\right)\notag\\
&\sim \frac{\sqrt{2\varR}}{\muR}\sqrt{\ln(\Psi)}.
\end{align}
This completes the proof. 

\vspace{-2mm}
\section{Proof of Proposition~\ref{Prop:2}}\label{App:proofoutGa}
With the Gaussian approximated PDF of $R_{\psi}$, we have the  approximated outage probabilities for the systems without and with reconfigurable antennas as
\begin{equation}\label{eq:outageGsin}
  \PP(R_{\psi}<R)=F_{R_\psi}(R)
\end{equation}
and
\begin{equation}\label{eq:outageGrec}
  \PP(R_{\cpsi}<R)=\left(F_{R_\psi}(R)\right)^{\Psi},
\end{equation}
respectively, where $F_{R_\psi}(x)$ is given in~\eqref{eq:CDFRpsi}.
Substituting \eqref{eq:outageGsin} and \eqref{eq:outageGrec} into \eqref{eq:outthsin} and \eqref{eq:outthrec}, respectively, we can obtain the approximated $R_{\cpsi}^{\out}$ and  $R_{\psi}^{\out}$ as
\begin{equation}\label{eq:derRcout}
  R_{\cpsi}^{\out}\approx F_{R_\psi}^{-1}(\epsilon^{\frac{1}{\Psi}})=\muR-\sqrt{2\varR}\mathrm{erf}^{-1}\left(1-2\epsilon^{\frac{1}{\Psi}}\right)
\end{equation}
and
\begin{equation}\label{eq:derRout}
R_{\psi}^{\out}\approx F_{R_\psi}^{-1}(\epsilon)=\muR-\sqrt{2\varR}\mathrm{erf}^{-1}\left(1-2\epsilon\right),
\end{equation}
respectively, where $F_{R_\psi}^{-1}(x)$ denotes the approximated inverse cdf of $R_{\psi}$.

Substituting \eqref{eq:derRcout} and \eqref{eq:derRout} into \eqref{eq:Defth_gain_out} completes the proof.

\section{Proof of Corollary~\ref{Cor:outGlsngrO}}\label{App:proofCor:outGlsngrO}
As $\Psi\rightarrow\infty$, we have $1-2\epsilon^{\frac{1}{\Psi}}\rightarrow-1$. From the tail region approximation for the inverse error function, 
we note that
\begin{equation}\label{eq:appinverflargexneg}
  \mathrm{erf}^{-1}(x)=  -\sqrt{-\ln\left(1-x^2\right)}  \quad \text{as} \quad x\rightarrow -1.
\end{equation}
Based on \eqref{eq:appinverflargexneg} and~\eqref{eq:GavelargePsi},
as $\Psi\rightarrow\infty$,  $G_{R^\out}(\Psi)$ is asymptotically equivalent to
\begin{align}
&G_{R^\out}(\Psi)\sim
\frac{\muR}{\muR-\sqrt{2\varR}\mathrm{erf}^{-1}\left(1-2\epsilon\right)}+\notag\\
&
\frac{\sqrt{2\varR}}{\muR-\sqrt{2\varR}\mathrm{erf}^{-1}\left(1-2\epsilon\right)}\sqrt{-\ln\left(1-\left(1-2\eps^{\frac{1}{\Psi}}\right)^2\right)}
\notag\\
&\sim \frac{\sqrt{2\varR}}{\muR-\sqrt{2\varR}\mathrm{erf}^{-1}\left(1-2\epsilon\right)}\sqrt{-\ln\left(1-\eps^{\frac{1}{\Psi}}\right)}
\notag\\
& \stackrel{(a)}{\sim}\frac{\sqrt{2\varR}}{\muR-\sqrt{2\varR}\mathrm{erf}^{-1}\left(1-2\epsilon\right)}\sqrt{\ln(\Psi)},
\end{align}
where $(a)$ is derived by analyzing the Taylor series of $\sqrt{-\ln\left(1-\eps^{\frac{1}{\Psi}}\right)}$ at $\Psi\rightarrow\infty$.
This completes the proof.

%


\balance

\end{document}